\documentclass{article}
\usepackage{spconf,amsmath,graphicx,hyperref}

\usepackage{amssymb,booktabs,pgfplots}
\usepackage{graphicx}

\usepackage{subcaption}
\usepackage{enumitem}
\pgfplotsset{compat=1.18}

\newcommand{\pavc}{P_{\mathrm{AVC}}}

\title{Perception-Inspired Bayesian Causal Fusion for\\Audiovisual Source Localization}
\name{Kyung Yun Lee$^{\sharp }$ \qquad Sungnyun Kim$^{\flat}$ \qquad Sebastian J. Schlecht$^{\natural}$ \qquad Tae-Hyun Oh$^{\flat}$ \qquad Vesa Välimäki$^{\sharp}$}
  
  \address{$^{\sharp}$ Acoustics Lab, Dept. of Information and Communications Eng., Aalto University, Espoo, Finland \\
      $^{\flat}$Korea Advanced Institute of Science and Technology (KAIST), Daejeon, South Korea\\$^{\natural}$Multimedia Comms. \& Signal Process., Friedrich-Alexander-Universität Erlangen-Nürnberg, Germany}
      
\begin{document}
\ninept
\maketitle
\begin{abstract}
Multimodal fusion promises more accurate perception but only when the modalities share a common cause. When they do not, the second modality carries no information about the target, and fusing it can only corrupt the estimate. We cast this \emph{whether-to-fuse} decision as Bayesian causal inference, following the optimal-observer model of human multisensory perception, and implement it as a plug-and-play layer on top of  frozen audio and visual models for sound event localization and detection. The model infers a common-cause posterior over visible candidates, then gates precision-weighted fusion accordingly. Fusing unconditionally more than doubles the direction error, whereas the causal gate improves on-screen localization while limiting off-screen degradation, without any joint network retraining.
\end{abstract}
%

% ========================================================================
\section{Introduction}
\label{sec:introduction}

% Humans are innately multimodal---we effortlessly combine sensory cues into a
% coherent understanding of the world, motivating a growing body of multimodal
% machine learning research~\cite{baltrusaitis2019,li2024multimodal}. Multimodal
% learning can outperform unimodal when modalities provide complementary
% information about a shared latent state~\cite{huang2021makes}---but the nature
% of this complementarity determines when fusion is beneficial and when it is
% harmful.
Humans combine sensory cues into a coherent understanding of the world, motivating a growing body of multimodal
machine learning research~\cite{baltruvsaitis2018multimodal,li2026multimodal}. Multimodal learning outperforms unimodal learning when the modalities carry complementary information about a shared latent state~\cite{huang2021makes}.
Therefore, whether this fusion helps or harms depends on whether such a shared state exists.

\begin{figure}[bt!]
    \centering

    \includegraphics[width=0.48\textwidth]{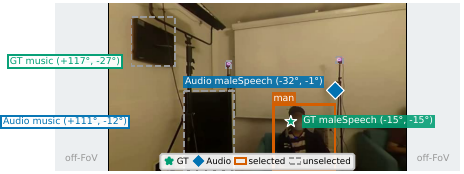}
    \hfill
    \includegraphics[width=0.48\textwidth]{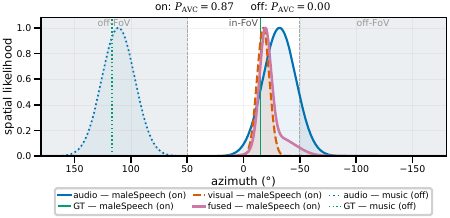}

    \caption{Overview of Bayesian causal fusion on an example frame. Top: video frame with detected objects (orange: selected candidate; grey: unselected), ground-truth source location, and audio-only estimate for two concurrent sound events---an on-screen event (\texttt{maleSpeech}) with a compatible visible candidate and an off-screen event (\texttt{music}) with no matching candidate. Bottom: audio and visual angular Gaussians for each modality and the resulting precision-weighted fused estimate. For the candidate identified as sharing a common cause, the fused estimate is closer to the ground-truth location than the audio-only estimate.}
    \label{fig:dog_conditions}
\end{figure}

Consider vision and hearing. Sound propagates around obstacles and arrives from every direction, but vision covers only a limited field of view. This asymmetry is natural for both humans and machines; a robot, a wearable device, or an AR headset likewise pairs a forward-facing camera with multi-channel microphones. A source that is occluded, outside the field of view, or behind the observer therefore remains audible but invisible (Figure~\ref{fig:dog_conditions}).
When a source \emph{is} visible,
% vision offers higher spatial precision than
% audio~\cite{kording2007,alais2004}, so combining them reduces
% error---but when it is not, the visual channel is conditionally
% independent of the source and fusing it injects an entirely wrong
% direction~\cite{juanola2025,tzinis2022dontlisten,audioscopev2}.
vision offers higher spatial precision than audio~\cite{kording2007causal,alais2004}, so combining them reduces error. When it is not visible, the visual channel is conditionally independent of the source, and fusing it pulls the estimate toward an unrelated entity~\cite{juanola2025,tzinis2022audioscopev2}.

Before asking \emph{how} to fuse, the observer must first determine \emph{whether} to fuse at all. Yet jointly trained networks delegate this decision to learned
representations~\cite{lu2023theory,berghi2024} with no explicit gate.
Uncertainty-aware methods weigh modalities by estimated
reliability~\cite{han2022trusted,shon2019noise,ovanger2026} but modulate influence
continuously rather than inferring whether a shared cause exists.
On-screen sound separation systems address a related trade-off by
learning to suppress off-screen audio through calibrated audiovisual
correspondence~\cite{tzinis2022audioscopev2,audioscope}, and
negative-audio evaluations expose the failure mode when this
distinction is ignored~\cite{juanola2025}. These methods require
end-to-end training on mixed on/off-screen data and target
source separation or localization heatmaps rather than spatial
event estimation.

Perception science offers a normative answer. K\"ording et
al.~\cite{kording2007causal} showed that human multisensory cue combination is well
described by an \emph{optimal observer} that performs \emph{causal inference}: the
brain first infers whether cues share a common cause, integrating them via
precision-weighted combination when they do and segregating when they do not. The ventriloquist effect and its
collapse at large disparities are natural consequences of this
framework~\cite{kording2007causal,alais2004,shams2010}. 
The entire benefit of integration depends on a shared cause, \textit{i.e.,} without one, the
second modality carries no information and integrating it can only degrade the
estimate.

We target a complementary setting: post-hoc gating of frozen
single-modality estimators for sound event localization and detection
(SELD)~\cite{adavanne2018}. We call this approach \emph{Bayesian causal fusion} (\textsc{Causal}), operationalizing
K\"ording et al.'s causal-inference observer~\cite{kording2007causal}
with three extensions:
(i)~generalizing the two-cue azimuth model to multiple visual
candidate rays on the sphere,
(ii)~operating post-hoc on frozen audio and visual models---no
component is retrained, and
(iii)~evaluating the resulting gate both as a common-cause classifier
and as a localization corrector under DCASE on/off-screen protocols.
We evaluate on the DCASE2025 Task~3 stereo on/off-screen
dataset~\cite{dcase2025} as well as a four-channel microphone
version~\cite{starss23} that we construct for full-sphere assessment.
Frozen audio~\cite{dcase2025, berg2024} and
visual~\cite{groundingdino} models serve as noisy estimators, much
like biological sensors, in a limited-FOV\,(field of view) camera
paired with spatial audio where the causal question arises naturally
at every frame.

The remainder of this paper is organized as follows. Section~\ref{sec:method} formulates the whether-to-fuse decision as Bayesian causal inference and derives the proposed multi-candidate fusion rule. Section~\ref{sec:experiments} evaluates the approach on two DCASE datasets in terms of common-cause classification and localization accuracy. Section~4 concludes the paper.

% ========================================================================
\section{Method}
\label{sec:method}

This section derives the Bayesian causal fusion model. We review the optimal-observer framework, define the audio and visual observations on the sphere, and present the proposed multi-candidate posterior and model-averaged integration rule in Sec.~\ref{sec:proposed}.

\subsection{Optimal observer model}

\begin{figure}[t]
    \centering
    \includegraphics[width=0.9\linewidth]{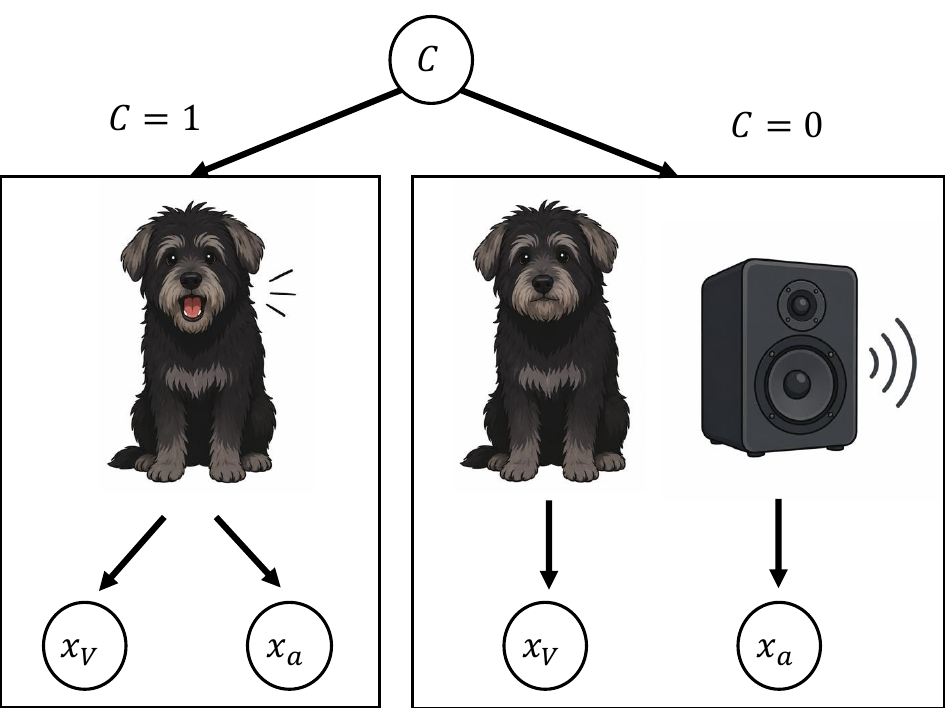}
    \caption{The two causal structures. $C{=}1$: a single source (the dog)
    generates both the visual observation $\mathbf{x}_v$ and the audio
    observation $\mathbf{x}_a$, so integration is beneficial. $C{=}0$: the
    visual and audio signals have independent causes, so integration corrupts
    the estimate.}
    \label{fig:causal_inference}
\end{figure}

Following K\"ording et al.~\cite{kording2007causal}, consider an observer that
must localize a source at true direction $\mathbf{s}\in\mathbb{S}^2$ from two
noisy sensory measurements: an audio observation
$\mathbf{x}_a$ and a visual observation $\mathbf{x}_v$. A latent causal variable $C$ determines the generative
structure (Figure~\ref{fig:causal_inference}):
\begin{itemize}[leftmargin=*,itemsep=1pt,topsep=2pt]
\item \textbf{Common cause} ($C=1$): a single source at direction
      $\mathbf{s}$ generates both observations $\mathbf{x}_a$ and
      $\mathbf{x}_v$, which are then noisy measurements of the same $\mathbf{s}$, so integrating them reduces localization uncertainty.
\item \textbf{Independent causes} ($C=0$): the audio and visual
      signals originate from different sources, \textit{i.e.,}
      $\mathbf{x}_v\perp\mathbf{s}\mid C{=}0$. The visual observation
      carries no information about the audio source direction, and the observer should segregate.
\end{itemize}
The optimal Bayesian observer must simultaneously infer the causal structure
and estimate source location. Under squared error, the optimal estimate
decomposes via the law of total expectation~\cite{kording2007causal}:
\begin{equation}
\hat{\mathbf{s}}^*
= p\,\hat{\mathbf{s}}_{\text{int}}
+ (1-p)\,\hat{\mathbf{s}}_{\text{seg}},
\label{eq:optimal}
\end{equation}
where $p=P(C{=}1\mid\mathbf{x}_a,\mathbf{x}_v)$ is the
common-cause posterior, $\hat{\mathbf{s}}_{\text{int}}$ is the precision-weighted integrated
estimate under assumed common cause, and $\hat{\mathbf{s}}_{\text{seg}}$ is
the segregated (audio-only) estimate under assumed independent causes.
Full integration ($p{=}1$) and full segregation ($p{=}0$) are special cases, and the optimal observer interpolates between them weighted by its belief about causal structure.
The common-cause posterior is the causal gate for multisensory
integration, which determines whether visual data has any information value at all.

\subsection{Audio and visual observations}

We now instantiate the abstract framework with concrete direction estimates:
the audio observation $\mathbf{x}_a$ becomes a decoded direction
$\mathbf{a}\in\mathbb{S}^2$, and each candidate visual observation becomes a
camera ray $\mathbf{v}_j\in\mathbb{S}^2$ indexed by detected object $j$. The
binary causal variable generalizes to multiple visual candidates,
$C\in\{0,1,\ldots,J_c\}$, where $C{=}j$ for $j{\geq}1$ means that candidate $j$ and the sound share a common source, and $C{=}0$ means that none of the visible candidates explains the sound, either because the source is off-screen, occluded, or missed by the detector.
% covering both off-screen sources and on-screen sources that are not identified due to occlusion or missed by object detector.

At each $100$-ms frame, a frozen audio model operating produces an event class $c$ and a Cartesian
unit direction $\mathbf a\in\mathbb S^2$.
A frozen open-vocabulary detector is prompted with labels of potentially
sounding objects (\textit{e.g.,} \emph{dog}, \emph{vacuum cleaner}, \emph{person}) and
produces labeled bounding boxes $\{b_j\}_{j=1}^{J}$, each with class
label~$\ell_j$. 
Because the detector and the audio model use different taxonomies,
a manual mapping pairs each of the 26 detector labels with a
compatible DCASE class
(\textit{e.g.,} \emph{vacuum cleaner}~$\to$~\emph{domestic sounds}); only detections
that match the current audio event~$c$ are retained as
candidates~$\mathcal{J}_c=\{j:\ell_j\in\mathcal{S}(c)\}$.
In principle, an LLM or audio-language embeddings such as
CLAP~\cite{elizalde2023clap} can advance this step~\cite{shimada2025open}.

Each bounding box center $(u_j, r_j)$ is projected to a unit camera ray with focal length $f=(W/2)\cot(\Omega/2)$, where $\Omega$ is the horizontal field of view:
\begin{equation}
 \mathbf v_j=
 \frac{[f,\ W/2-u_j,\ H/2-r_j]^\top}
 {\|[f,\ W/2-u_j,\ H/2-r_j]^\top\|_2}\in\mathbb S^2,
 \label{eq:visual_ray}
\end{equation}
where $W/2{-}u_j$ and $H/2{-}r_j$ are the pixel offsets from the image center along the azimuth and elevation axes, respectively.

\subsection{Angular likelihood}

Audio and visual localization uncertainty is estimated on training events as
great-circle RMS errors $\sigma_a$ and $\sigma_v$. Their common-cause
separation has RMS
$\sigma_\Delta=(\sigma_a^2+\sigma_v^2)^{1/2}$, expressed in radians below. We
model a candidate direction under a common cause by a von Mises--Fisher (vMF)
density centered on $\mathbf a$,
\begin{equation}
 p(\mathbf v_j\mid\mathbf a,C=j)
 =\frac{\kappa\exp(\kappa\mathbf a^\top\mathbf v_j)}
 {4\pi\sinh\kappa},\quad
 \kappa=\frac{2}{\sigma_\Delta^2}.
 \label{eq:vmf}
\end{equation}
Under $C{=}0$ (no common cause), a candidate direction is uniform over solid angle,
$p(\mathbf v_j\mid C=0)=1/(4\pi)$. The spherical angular Bayes
factor is therefore
\begin{equation}
 \Lambda_j
 =\frac{\kappa}{\sinh\kappa}
   \exp(\kappa\mathbf a^\top\mathbf v_j).
 \label{eq:angular_bf}
\end{equation}

\subsection{Causal posterior and fusion}
\label{sec:proposed}

Let $J_c=|\mathcal{J}_c|$ and let $\pi_{\rm vis}$ denote the training frequency
of on-screen events among candidate-eligible samples, used as a proxy prior for
a visible cause. This prior is divided equally among candidates:
$P(C{=}j)=\pi_{\rm vis}/J_c$,
$P(C{=}0)=1-\pi_{\rm vis}$.
The common-cause posterior is
\begin{equation}
 P\!\left(C=j\mid\mathbf a,\mathbf v_{1:J},c\right)
 =\frac{(\pi_{\rm vis}/J_c)\Lambda_j}
 {(1-\pi_{\rm vis})+
  \sum_{k\in\mathcal{J}_c}(\pi_{\rm vis}/J_c)\Lambda_k}.
 \label{eq:posterior}
\end{equation}
For brevity, let $p_j\triangleq P(C{=}j\mid\mathbf a,\mathbf v_{1:J},c)$
denote this posterior. We use $\pavc=1-p_0$ as the common-cause
posterior, the multi-candidate generalization of $p$ in
\eqref{eq:optimal}. When $J_c{=}0$, the independent-cause probability is 1.

Implementing the causal gate of \eqref{eq:optimal}, we first infer
causal structure, then integrate. Conditioned on $C{=}j$, the precision-weighted integrated direction estimate is
\begin{equation}
 \hat{\mathbf s}_j=
 \frac{\sigma_a^{-2}\mathbf a+\sigma_v^{-2}\mathbf v_j}
 {\|\sigma_a^{-2}\mathbf a+\sigma_v^{-2}\mathbf v_j\|_2}.
 \label{eq:fused_location}
\end{equation}
We propose to obtain the final estimate by directly implementing
\eqref{eq:optimal} as a multi-candidate model average on the sphere:
\begin{equation}
 \hat{\mathbf s}=
 \frac{p_0\,\mathbf a+\sum_{j\in\mathcal{J}_c}p_j\,\hat{\mathbf s}_j}
      {\|p_0\,\mathbf a+\sum_{j\in\mathcal{J}_c}p_j\,\hat{\mathbf s}_j\|_2}.
 \label{eq:model_average}
\end{equation}
The denominator projects the weighted combination back to the unit sphere.
When $\pavc\approx 0$, the visual terms vanish and the estimate reduces to the
audio direction~$\mathbf a$; when $\pavc\approx 1$, it reduces to
precision-weighted cue combination with the dominant candidate.
Averaging over all candidates rather than selecting the most probable one
($\arg\max_j p_j$) avoids discontinuities: when two candidates carry
similar posterior mass, a small change in the audio observation can flip
the winner, causing the fused estimate to jump. The model average moves
continuously as posterior mass shifts between candidates.
% No threshold
% is needed: the posterior continuously gates how much vision influences the
% estimate, implementing the conditional independence of
% Sec.~\ref{sec:method}---when the posterior favors $C{=}0$, the
% estimate automatically ignores a visual channel that carries no information.
No threshold is needed. The posterior continuously gates how much vision influences the estimate, reflecting the conditional independence $\mathbf{x}_v\perp\mathbf{s}\mid C{=}0$.

% \begin{figure}[t]
% \centering
% \includegraphics[width=0.98\linewidth]{figures/causal_curves.pdf}
% \caption{Illustrative full-sphere angle-only response. (a) The common-cause
% posterior falls with spatial disparity, closing the information gate.
% (b) Consequently, the causal inference model's bias toward vision first grows
% and then recedes as the observer transitions from integration to segregation;
% full integration has no such gate.}
% \label{fig:causal_behavior}
% \end{figure}

% ========================================================================
\section{Experiments and Results}
\label{sec:experiments}

We evaluate on two dataset configurations, assessing common-cause classification (Sec.~\ref{sec:results_congruence}), localization accuracy (Sec.~\ref{sec:results_gate}), and comparison with the jointly trained DCASE2025 baseline (Sec.~\ref{sec:stereo}).

\subsection{Setup}
\label{subsec:setup}

\textbf{Datasets.}~~
We evaluate on two versions of the DCASE2025 Task~3
dataset~\cite{dcase2025}, both derived from STARSS23~\cite{starss23}.
\emph{(i)~Stereo}: the original DCASE2025 release with stereo audio (azimuth only) and perspective video at
$640{\times}360$ with horizontal FOV $\Omega{=}100^\circ$~\cite{dcase2025}.
\emph{(ii)~Four-channel microphone (MIC)}: a tetrahedral array
counterpart using the same temporal crops and camera yaws, providing
full-sphere (azimuth + elevation) evaluation. The microphone directions
are rotated to the camera viewing direction so that audio and visual
rays share the same coordinate frame. Multi-channel arrays are
standard for spatial scene understanding~\cite{starss23}, making this
a more realistic setting.

No ground-truth common-cause label exists, so we evaluate the common-cause posterior
against two proxies.
\emph{(a)~On-screen}: the DCASE2025 human annotation of whether the source is within the field of view~\cite{dcase2025}. This is a necessary precondition for a visual common cause, but permissive: a source may be on-screen without a matching detection. 
\emph{(b)~Common-cause}: an on-screen event whose ground-truth location lies within $20^\circ$ of a semantically compatible Grounding DINO detection. This is a tighter approximation of whether a matched visual candidate exists, but it shares the detector and semantic mapping with the causal gate's candidate set, so high agreement is partly a consequence of shared inputs rather than independent validation of causal inference.
% \emph{(a)~On-screen} (DCASE ground truth)---necessary for a visual
% common cause but permissive, since a source may be on-screen without a
% matching detection;
% \emph{(b)~common-cause}---for a sounding object labeled on-screen by
% the DCASE2025 ground truth, if a semantically compatible Grounding
% DINO detection falls within $20^\circ$ of the ground-truth location,
% the event is labeled as common-cause.

\noindent\textbf{Models.}~~
For stereo, we use the official DCASE2025
checkpoints\footnote{\href{https://github.com/partha2409/DCASE2025_seld_baseline}{https://github.com/partha2409/DCASE2025\_seld\_baseline}}:
% \footnote{\url{https://github.com/partha2409/DCASE2025_seld_baseline}}:
audio-only SELDnet (\textsc{Audio}) and jointly trained audiovisual
SELDnet (\textsc{AV-base}).
For microphone, we use an available open source model from the DCASE2024 challenge~\cite{berg2024}\footnote{\href{https://github.com/axeber01/ngcc-seld}{https://github.com/axeber01/ngcc-seld}}.
In both settings, Grounding DINO Tiny~\cite{groundingdino} serves as
the open-vocabulary detector, prompted with a fixed vocabulary of
26~sounding-object labels.
The semantic mapping $\mathcal{S}(c)$ (Sec.~\ref{sec:method}) pairs
each detector label with one of the 13 DCASE classes.
All components are frozen, and causal fusion is applied post-hoc as described in
Sec.~\ref{sec:method}.

\subsection{Compared methods}
\label{sec:methods}

Audio and visual detections are fixed across all methods; only the fusion
strategy varies. We compare four strategies
that differ in how/whether the common-cause posterior $\pavc$ gates
fusion, with the audio-only estimate being the baseline and the last being the method from Sec.~\ref{sec:proposed}:
\begin{itemize}[leftmargin=*,itemsep=1pt,topsep=1pt]
\item \textsc{Audio}: retain audio direction ($\pavc{=}0$; full
      segregation).
\item \textsc{Forced}: always fuse with the nearest compatible candidate
      ($\pavc{=}1$; full integration).
\item \textsc{Gate-20}: fuse only when audio--visual disparity $<20^\circ$
      (hard angular gate).
\item \textsc{Causal} (Bayesian causal fusion; the proposed method): infer $\pavc$ via \eqref{eq:posterior}, then
      model-average via \eqref{eq:model_average}.
\end{itemize}
\textsc{Audio} and \textsc{Forced} are the two limiting cases of the causal
inference model~\cite{kording2007causal}. \textsc{Gate-20} is an ad-hoc threshold.
\textsc{Causal} tests whether the normative solution provides a better gate.

\subsection{Common-cause classification}
\label{sec:results_congruence}

\begin{table}[!t]
\centering
\caption{Common-cause classification on eligible events using two proxy
labels. Higher AUROC and AP indicate better discrimination; lower Brier score
and ECE indicate better calibration. Calibration metrics are reported only for
the probabilistic \textsc{Causal} posterior.}
\label{tab:congruence}
\scriptsize
\setlength{\tabcolsep}{2.5pt}
\begin{tabular}{llccrr}
\toprule
Method & Proxy & AUROC & AP & Brier & ECE\\
\midrule
\textsc{Forced}  & on-screen & 0.500 & 0.332 & --- & ---\\
\textsc{Gate-20} & on-screen & 0.828 & 0.723 & --- & ---\\
\textsc{Causal} (proposed)    & on-screen & \textbf{0.920} & \textbf{0.847} & 0.093 & 0.060\\
\midrule
\textsc{Forced}  & common-cause & 0.500 & 0.302 & --- & ---\\
\textsc{Gate-20} & common-cause & 0.839 & 0.706 & --- & ---\\
\textsc{Causal} (proposed)     & common-cause & \textbf{0.924} & \textbf{0.805} & 0.085 & 0.048\\
\bottomrule
\end{tabular}
\end{table}

Before evaluating localization, we ask whether the inferred common-cause
posterior discriminates events for which visual integration is plausible
(Table~\ref{tab:congruence}), using the two proxies from Sec.~\ref{subsec:setup}.

An event is \emph{eligible} when the detector finds at least one compatible
object; otherwise all methods reduce to the audio-only estimate.
Of the eligible events, only 33.2\% are on-screen,
while 66.8\% are off-screen. The stricter proxy identifies
30.2\% as plausible common-cause cases.

The detector-independent on-screen label provides the stronger test.
The continuous posterior discriminates on-screen from off-screen events
(AUROC $0.920$, AP $0.847$; Table~\ref{tab:congruence}), substantially
exceeding chance. Because \textsc{Gate-20} is a binary decision, its AUROC
reduces to balanced accuracy at a single operating point; the localization
comparison in Sec.~\ref{sec:results_gate} provides the primary evaluation.
Against the common-cause proxy, \textsc{Causal} achieves AUROC $0.924$ and
AP $0.805$. This second result is less independent because the
proxy reuses the detector and semantic mapping that construct the model's
candidate set. Neither proxy directly annotates causal structure, so these
results demonstrate the gate's discriminative value rather than recovery of a
latent causal variable.

Because the posterior is used as a continuous fusion weight, discrimination
alone is insufficient: its probabilities must also be calibrated. We measure
calibration with the Brier score (mean squared error between predicted
probability and label) and expected calibration error (ECE; mean absolute
gap between predicted confidence and observed accuracy across probability
bins). \textsc{Causal} obtains a Brier score of $0.093$ and an ECE of $0.060$
against the on-screen proxy. Against the common-cause proxy, the corresponding values are $0.085$
and $0.048$. Thus, within a semantically compatible candidate set, spatial
disparity alone yields a discriminative and well-calibrated gate with respect
to these proxies, without spectral, temporal, or learned cross-modal features.

For reproducibility, training-split estimation yielded
$\sigma_a{=}14.7^\circ$, $\sigma_v{=}5.5^\circ$, and
$\pi_{\rm vis}{=}0.33$; all three values were frozen before test-set
evaluation.

% \begin{figure}[t]
%     \centering
%     \input{reliability_diagram}
%     \caption{Reliability diagram of $\pavc$ on eligible events against the
%     on-screen proxy. Dot area is proportional to the log event count; the
%     dashed line is perfect calibration.}
%     \label{fig:reliability}
% \end{figure}

\subsection{Localization}
\label{sec:results_gate}

% [Old event-weighted version kept for reference]
% \textsc{Forced} fusion---always integrating---improves on-screen DOAE by
% $9.16^\circ$ but increases off-screen DOAE by $72.88^\circ$, more than
% tripling eligible-event error from $20.00^\circ$ to $65.64^\circ$. Across
% all $603{,}152$ test events, forced fusion more than doubles overall DOAE
% from $21.22^\circ$ to $49.49^\circ$.

Table~\ref{tab:causal} reports the effect on localization. Since only a minority of eligible events are on-screen, overall gains are modest even when per-condition effects are large.
\textsc{Forced} fusion improves on-screen DOAE but roughly doubles eligible-event error, because off-screen events are pulled toward unrelated detections. This is the cost of integrating when $C{=}0$.

\begin{table}[t]
\centering
\caption{MIC full-sphere class-macro DOAE ($^\circ$) on eligible events, following the official DCASE protocol. $\Delta$ ($^\circ$) is relative to \textsc{Audio}. Lower is better.}
\label{tab:causal}
\scriptsize
\begin{tabular}{lcccrr}
\toprule
&\multicolumn{3}{c}{DOAE ($^\circ$)}&\multicolumn{2}{c}{$\Delta$ ($^\circ$) from audio}\\
\cmidrule(lr){2-4}\cmidrule(lr){5-6}
Method & All & On & Off & On & Off\\
\midrule
\textsc{Audio}      & 30.41 & 28.15 & 30.95 &  0.00 &  0.00\\
\textsc{Forced}     & 61.47 & 12.74 & 89.20 & $-$15.41 & +58.26\\
\midrule
\textsc{Gate-20}    & 29.64 & 26.21 & 31.03 & $-$1.94 & +0.08\\
\textsc{Causal} (proposed)       & \textbf{29.50} & \textbf{25.40} & 31.20 & $-$\textbf{2.75} & +0.26\\
\bottomrule
\end{tabular}
\end{table}

Table~\ref{tab:causal} also reveals that \textsc{Causal} improves on-screen
DOAE by $2.75^\circ$ while limiting off-screen degradation to $+0.26^\circ$,
reducing eligible-event error by $0.91^\circ$.
% BCI achieves a larger on-screen gain ($3.98^\circ$ vs.\ $2.75^\circ$) because
% the probabilistic gate opens more confidently for strong candidates rather
% than applying a single fixed threshold. The asymmetry---$+72.88^\circ$
% off-screen harm without the gate versus $+0.36^\circ$ with it---is the
% central result: no integration mechanism can extract useful information from a
% channel that carries none; the causal gate prevents the attempt.
Its on-screen gain exceeds that of \textsc{Gate-20} ($2.75^\circ$ vs.\ $1.94^\circ$) because the posterior weights strong candidates more heavily instead of applying one fixed threshold.

\subsection{Comparison with DCASE 2025 Task 3 baseline}
\label{sec:stereo}

The DCASE 2025 Task~3 baseline includes a jointly trained audiovisual model, enabling a direct comparison with a learned fusion approach. We apply causal fusion to the stereo dataset using the official audio-only checkpoint~\cite{dcase2025}.
Since stereo front-back ambiguity folds azimuths into
$[-90^\circ,90^\circ]$, the vMF likelihood reduces to a wrapped
Gaussian; the model is otherwise identical to Sec.~\ref{sec:method}.
We compare against both official baselines (\textsc{Audio} and the
jointly trained \textsc{AV-base} \cite{dcase2025}).
On/off-screen classification accuracy is defined as
$\mathbf{1}[\pavc > 0.5]$ for \textsc{Causal} and the model's internal flag for the \textsc{AV-base}.

Table~\ref{tab:stereo} shows that \textsc{Causal} reduces on-screen DOAE by $2.81^\circ$ relative to \textsc{Audio}.
The jointly trained \textsc{AV-base} achieves the lowest aggregate and off-screen DOAE, but its on-screen DOAE ($19.13^\circ$) is worse than both \textsc{Audio} ($16.59^\circ$) and \textsc{Causal} ($13.78^\circ$), indicating that joint training does not learn an effective on/off-screen gate.
\textsc{Causal} attains higher on/off-screen accuracy ($0.76$ vs.\ $0.74$) without any joint training.

\begin{table}[t]
\centering
\caption{DCASE2025 stereo comparison on the test split (class-macro DOAE, following the official protocol). \textsc{Audio} and
\textsc{AV-base} are the two official DCASE2025 baseline
checkpoints~\cite{dcase2025}; \textsc{Causal} applies post-hoc causal
fusion to \textsc{Audio}'s detections. DOAE ($^\circ$, lower is
better); on/off-screen accuracy (higher is better).}
\label{tab:stereo}
\scriptsize
\begin{tabular}{lcccr}
\toprule
&\multicolumn{3}{c}{DOAE ($^\circ$)}&{On/Off}\\
\cmidrule(lr){2-4}
Method & All & On & Off & Accuracy\\
\midrule
\textsc{Audio}     & 24.58 & 16.59 & 25.74 & ---\\
\textsc{AV-base}   & \textbf{22.16} & 19.13 & \textbf{22.89} & 0.74\\
\midrule
\textsc{Causal} (proposed) & 24.24 & \textbf{13.78} & 25.96 & \textbf{0.76}\\
\bottomrule
\end{tabular}
\end{table}

% ========================================================================
\section{Conclusion}

% We brought Bayesian causal inference from perception
% science~\cite{kording2007} into audiovisual machine learning, casting the
% \emph{whether-to-fuse} decision as an explicit common-cause posterior that
% gates multisensory integration. The gate is necessary: without it, fusion
% with off-screen sources is catastrophic because the visual channel is
% conditionally independent of the source. The gate is sufficient: once causal
% structure is correctly inferred, a simple precision-weighted cue combination
% captures the on-screen benefit while containing off-screen harm. Beyond this
% specific task, we believe the framework offers a design principle for
% multimodal systems more broadly. Rather than delegating the integration
% decision to end-to-end learned representations, systems can first infer
% whether modalities share a common cause and only then combine them. A further
% advantage is that the Bayesian formulation naturally models
% uncertainty---inherent in all perception and estimation problems yet often
% difficult to incorporate in deep learning---and uses it directly to make the
% fusion decision. The question for multimodal systems is not \emph{how} to
% integrate, but \emph{whether}.

We cast the \emph{whether-to-fuse} decision as Bayesian causal inference, following the optimal-observer model of human multisensory perception~\cite{kording2007causal}, and applied it as a plug-and-play layer above frozen audio and visual models for sound event localization and detection.
Without the gate, fusion with off-screen sources drastically raises direction error, because the visual channel is conditionally independent of the source. With the gate, precision-weighted cue combination captures the on-screen benefit while containing off-screen harm.
Because the gate operates post-hoc on frozen models, it is plug-and-play: no multimodal training is needed, and any improved audio or visual model can be substituted directly. On two datasets, this zero-training-cost fusion improves on-screen DOAE over the audio-only baseline. On the stereo dataset, it also achieves better on-screen localization and on/off-screen classification than a jointly trained audiovisual system~\cite{dcase2025}.
The gate currently relies on an open-vocabulary detector and a manual binary class mapping; failures in either reduce the candidate set and default to audio-only, which is safe but forgoes visual benefit. A natural extension is to replace the hard semantic filter with a soft compatibility score from an audio-language model such as CLAP~\cite{elizalde2023clap}, removing the need for a manual class mapping.
The same principle of inferring whether modalities share a common cause before combining them may apply to other multimodal settings where the second modality is not always informative.

\section{ACKNOWLEDGMENTS}
This work was supported by the HUCE infrastructure of the Aalto School of Electrical Engineering. SJS were supported through the joint German Academic Exchange Service (DAAD) and Research Council of Finland (RCF) Project (57763119) "Immersive Augmented Acoustics (IAA)".

% \section{REFERENCES}
% \label{sec:refs}

\bibliographystyle{IEEEbib}
\bibliography{strings,refs}

\end{document}